\documentclass[11pt,letterpaper,oneside]{article}
\usepackage[bindingoffset=0in, left=1in, right=1in, top=1in, bottom=1in]{geometry}
\usepackage{graphicx}
\usepackage{authblk}
\usepackage{color}
\usepackage{lineno}
\usepackage{qcircuit}   
\usepackage{physics}    
\usepackage{url}

\newcommand{\coolname}{\texttt{qec\_code\_sim}}

\title{\coolname: An open-source Python framework for estimating the effectiveness of quantum-error correcting codes on superconducting qubits}
\author[1]{Santiago Lopez}
\author[2]{Jonathan Andrade Plascencia}
\author[3]{Gabriel N. Perdue}
\date{August 2023}

\affil[1]{Purdue University}
\affil[2]{Cornell University}
\affil[3]{Fermi National Accelerator Laboratory}

\newcommand{\reportnumber}{FERMILAB-PUB-24-0035-ETD}
\usepackage[angle=0,scale=1,color=black,firstpage=true,opacity=1]{background}

\SetBgContents{\small\texttt{\reportnumber}}
\SetBgPosition{current page.north east}
\SetBgHshift{-6.5cm}
\SetBgVshift{-1.5cm}

\begin{document}
\maketitle

\begin{abstract}
Quantum computers are highly susceptible to errors due to unintended interactions with their environment.
It is crucial to correct these errors without gaining information about the quantum state, which would result in its destruction through back-action.
Quantum Error Correction (QEC) provides information about occurred errors without compromising the quantum state of the system.
However, the implementation of QEC has proven to be challenging due to the current performance levels of qubits --- break-even requires fabrication and operation quality that is beyond the state-of-the-art.
Understanding how qubit performance factors into the success of a QEC code is a valuable exercise for tracking progress towards fault-tolerant quantum computing.

Here we present \coolname, an open-source, lightweight Python framework for studying the performance of small quantum error correcting codes under the influence of a realistic error model appropriate for superconducting transmon qubits, with the goal of enabling useful hardware studies and experiments.
\coolname requires minimal software dependencies and prioritizes ease of use, ease of change, and pedagogy over execution speed.
As such, it is a tool well-suited to small teams studying systems on the order of one dozen qubits.
\end{abstract}


\section{\label{sec:intro}Introduction}


Quantum computers use the properties of quantum states, like superposition and entanglement, to run certain algorithms with the potential for speed-ups with polynomial, or even exponential, scaling in problem size over classical computers \cite{Maslov2019}.
Quantum computers operate on states stored in a real or artificial atom, or some other kind of quantum device.
The basic unit of computation is usually a two-level quantum system referred to as a qubit. 
These computers may hopefully address many significant problems in industry and basic research: logtistics, finance, cryptography, materials science, chemistry, optimization, quantum gravity, particle physics, and non-equilibrium many-body dynamics among others \cite{Altman2021,Bova2021}.
There are many different physical architectures to choose from when building a quantum computer. 
Some use hundreds of Rydberg atoms trapped in optical tweezers \cite{Scholl2021}, superconducting Josephson junctions, or ultracold molecules. 

Although theoretically powerful, quantum computers are limited in practice by continuously growing errors.
The source of a quantum computer's power --- leveraging superposition and entanglement --- also prevents us from using many of the standard tools for error correction from information theory \cite{10.21468/SciPostPhysLectNotes.70}.
Quantum error correction (QEC) is the attempt to remove noise and errors by encoding multiple physical qubits into ``logical'' qubits and performing operations on the logical qubits that identify error syndromes while preserving the encoded logical state \cite{Shor1995}.
Upon first inspection, this strategy sounds hopeless given that any attempts to repair errors in a faulty quantum circuit necessarily require the introduction of additional layers of faulty quantum circuits which are themselves subject to quantum noise and errors.
Fortunately, some codes theoretically allow for \emph{fault-tolerant} quantum computing (FTQC) \cite{10.21468/SciPostPhysLectNotes.70} in which faulty quantum circuits may be used to correct errors in a computer with exponentially improving performance in the depth of the error-correcting hardware.
For fault-tolerant quantum computing to work, the intrinsic error rates in individual physical qubits must be below certain thresholds.
However, accurately characterizing these thresholds is not simple due to the many-faceted natures of quantum noise and qubit performance.

The cross-over point is referred to as ``break-even'' where we can fix errors in the circuit faster than the error correction circuit introduces them.
At the time of writing the record for a break-even extension is a factor of 2.3 \cite{Sivak2023}.
The general break-even problem is extremely difficult.
The characteristics of noise and errors in qubits are complex and qubit designers are often forced to make trade-offs and optimize for better performance in the case of some error sources as opposed to others.
Therefore, it is necessary on the path to fault tolerance to build an understanding of how different sources of error interact and influence the overall performance of a logical qubit.

To support this effort, we created \coolname: an open-source, Python-based software package to effectively emulate various QEC protocols against realistic error models, with a current focus on superconducting transmon qubits.
This package enables research groups to test quantum circuits with up to $\sim$12 qubits, which is a desktop-friendly scale, using specific noise model parameters that match their own physical devices.
Additionally, we included an introductory ``knowledge base'' of pedagogical material, making the information accessible to a broad audience.
The primary emphasis in the design of \coolname was to make the software portable, easy to extend, and easy to learn from, with less emphasis on execution speed.
This project is still in progress and available for collaboration and contribution on GitHub \footnote{\url{https://github.com/Lopez-Santi/qec_code_sim}}.

\subsection{Industrial and Research Applications}

QEC is widely expected to be a pre-requisite for major commercial and scientific impact.
By focusing on easily modifiable code, portability, pedagogical support, and a completely open license, \coolname will be an effective tool for small research teams and start-up companies focused on qubit fabrication and characterization.

\section{Prior Work}
\label{sec:priorwork}

Currently, there are several platforms for simulating quantum error correction codes including \cite{qecsim, Huang_2023, Priya_2018}. These platforms are focused on aspects including speed, visualization, code complexity, and decoding rather than pedagogy and ease of use. Furthermore, surface codes are a large focus in recent research for simulating FTQC \cite{Katsuda_2024} as these sets of codes can correct relaxation errors. Although surface codes require a large number of qubits and thus high computational resources, groups have implemented sparse linear algebra and graphical methods \cite{MacKay_2004, chancellor_graphical_2023} to help with computing time and memory. Many software packages help with circuit analysis and simulation for specific stabilizer codes, surface codes, and even user-made codes \cite{Qiskit}.

In addition to circuit simulation, some work has been done to turn these logical circuits into physical implementations for different qubit architectures \cite{Childs_2019, paler_mapping_2014}. This will prove valuable as QEC has been mainly based on theoretical approaches, but now experimentalists can test QEC codes on real hardware \cite{Sivak2023}.

Although \coolname does not prioritize execution time or code complexity, the software package enables the user to learn basic QEC protocols that are the foundation of more complex codes. Additionally, small groups wanting to test current superconducting qubit parameters against realistic error models can do this using \coolname.

\section{\label{sec:software}Software dependencies and organization}

The \coolname software is programmed using only Python Programming Language and the core functionality relies on the \texttt{NumPy} library \cite{harris2020array} heavily for mathematical simulation.
The optional user interface tool makes use of \texttt{h5py} \cite{h5py_2014} and tabulate for saving and managing data files.
Additionally, the optional pedagogical notebooks included with the library utilize \texttt{matplotlib} for plotting \cite{Hunter2007}, \texttt{SciPy}for fitting various curves and some minor math \cite{2020SciPy-NMeth}, \texttt{Qiskit} and \texttt{qiskit-aer} and \texttt{pylatexenc} for drawing some of the circuits used \cite{Qiskit}, and \texttt{prettytable} \cite{prettytable} for outputting information about a QEC circuit in a useful manner.

The current structure of the \coolname software is organized to be utilized in two ways.
The first is to test the performance of a QEC algorithm in maintaining a quantum state for one logical qubit given a set of values for the parameterized error model.
A user-friendly notebook, \texttt{QEC Simulator.ipynb}, is provided to illustrate this use-case. 
After inputting the parameters of a qubit system, different information about the system including the logical T1, the distribution of times at which your circuit will fail, and the distribution of the estimated logical T1 of your system will be output.

The second way to use \coolname is as an introduction to QEC.
There are notebooks inside the folder \emph{Implementation Knowledge Base} which include many introductory topics in QEC. 
These notebooks largely follow the review article \cite{Devitt_2013}, and go through the basics of QEC, filling in the material found there with detailed calculations. 
Through these pedagogical notebooks we introduce the concept of quantum errors, give many examples and implementations of different types of QEC algorithms, explain the full error model, and explain and implement fault tolerance.
Additionally, these files walk through how the \texttt{QEC Simulator.ipynb} works. 

\coolname currently supports four different QEC algorithms: the 3-qubit code, the 7-qubit Steane code, the fault-tolerant 7-qubit Steane code, and the 9-qubit code.
These codes are tested against realistic error models that will be discussed in Section \ref{sec:errors}. 

\section{\label{sec:errors}Quantum errors in superconducting qubits}


\coolname currently models three types of continuous errors: depolarization, relaxation and dephasing, and state preparation and measurement (SPAM). 
In this work, we closely followed \cite{PhysRevA.104.062432} and notes by John Preskill \cite{Preskill} when implementing these error channels.
We will further describe how our error model works for each type of error in Sections \ref{sec:depolar} - \ref{sec:spam}.

Generally, each of these errors is continuously occurring on the qubit system, causing the circuit failure over time. Although errors are continuous, in the simulation, qubit errors are implemented as discrete matrix operations. Since each gate operation in a real quantum circuit takes time to execute we took this into account to calculate the error matrix applied. After every circuit gate, one error gate is applied for each of the errors above (except SPAM since those occur once at the beginning and end of the circuit). This gives a realistic model for how these errors affect a QEC circuit over time. Although the model used is believed to be accurate, there are other errors, such as cross-talk, that have not been implemented.

\subsection{\label{sec:depolar}Depolarization Error Model}
The depolarization errors consist of random $X, Y, Z$ rotations caused by the intrinsic coupling of the qubit to the environment when applying gate operations. In our model, taken from \cite{PhysRevA.104.062432}, 4 different gates are applied to the density matrix, $\rho$, based on a probability, $p_1$, of them occurring. This probability is an estimate of a coupling parameter for each individual qubit in the system, thus if there is a low chance of error, this probability will be low. The 4 operators below were applied to each qubit after every gate operation in the QEC circuit.

$$K_{D_0} = \sqrt{1-p_1}, \quad K_{D_1} = \sqrt{\frac{p_1}{3}}X, \quad K_{D_2} = \sqrt{\frac{p_1}{3}}Z, \quad K_{D_3} = \sqrt{\frac{p_1}{3}}Y$$

$$\rho \mapsto D(\rho) = \sum_{i=0}^{3}K_{D_i}\rho K_{D_i}^{\dagger}$$

When performing a single qubit gate operation, the density matrix, $\rho$, is mapped to $D(\rho)$ for the qubit we are performing the gate on. However, when a 2 qubit gate is performed, such as a $CNOT$ gate, only the target qubit has a depolarization error applied to it. In this case, since the control qubit is not being manipulated there is no need to apply a depolarizing gate error on it. Lastly, \coolname offers a qubit-by-qubit variation in depolarization error probability. With these ideas in mind, the depolarization error model is a realistic method for simulating quantum operations.

\subsection{\label{sec:relaxdeph}Relaxation and Dephasing Error Models}
Qubits experience thermal decoherence (relaxation) and dephasing over time depending on the temperature of the environment. In our error model, both of these error channels are included to ensure realistic results. When implementing these time-dependent operations, it was best to do them discretely after each circuit gate operation. Estimating the time for each quantum gate, $T_g$, allows an accurate model for time-dependent errors.

\subsubsection{\label{sec:relax} Thermal Decoherence (Relaxation)}
Thermal decoherence is the energy exchange between the qubit and the environment along the x and y-axis of the bloch sphere, and it is irreversible. The characteristic time it takes a qubit to relax is known as $T_1$ and is defined by the evolution towards the equilibrium state at the temperature of the environment. The probability for each qubit to relax to the ground state is given by $P_{T_1}(q) = e^{-T_g/T_1(q)}$, where $(q)$ is the index of the qubit. More details on how this was derived can be found in \cite{PhysRevA.104.062432}. From this result, the probability of resetting to an equilibrium state is given by $p_{reset} = 1 - P_{T_1}$. Two operators are used when modeling this type of error, one acts as an identity operation and the other acts as a relaxation.
$$K_0 = \begin{pmatrix} 1 & 0 \\ 0 & \sqrt{1-p_{reset}} \end{pmatrix}, \quad K_1 = \begin{pmatrix} 0 & \sqrt{p_{reset}} \\ 0 & 0 \end{pmatrix}$$

\subsubsection{\label{sec:deph} Dephasing}
Dephasing is the transition of a quantum state to a classical one and is also due to the environmental coupling. This error occurs along the z-axis of the bloch sphere. Similarly to relaxation, the characteristic time it takes a qubit to dephase is known as $T_2$ and is defined by the behavior of off-diagonal entries in the density matrix over time.  The probability for each qubit to depahse is given by $P_{T_2}(q) = e^{-T_g/T_2(q)}$, where $(q)$ is the index of the qubit. More details on how this was derived can be found in \cite{PhysRevA.104.062432}. From this result, the probability of dephasing (phase flip) is given by $p_{dephase} = 1 - P_{T_2}$. Three operators are used when modeling this type of error, one acts as an identity operation and the other 2 act as a phase change.
$$K_2 = \sqrt{1-p_{dephase}}I, \quad K_3 = \sqrt{p_{dephase}}\vert0\rangle\langle0\vert, \quad K_4 = \sqrt{p_{dephase}}\vert1\rangle\langle1\vert$$

Therefore, when combining both Relaxation and Dephasing error models, the total effect on the density matrix, (of a single qubit) $\rho_q$ can be found, where $q$ is the qubit index. The error operators were applied after each gate operation to every qubit in the system. Lastly, although it can be done similarly to the depolarization error model, \coolname does not currently offer a qubit-by-qubit variation on relaxation and dephasing error rates.

$$\rho_q \mapsto \eta(\rho_q) = \sum_{k=0}^{4}K_{k}\rho_q K_{k}^{\dagger}$$

In further sections we will also discuss the gate time that we selected, as well as the $T_1$ and $T_2$ times of each of the qubits.

\subsection{\label{sec:spam}SPAM Error Models}
The State Preparation and Measurement (SPAM) Errors can be treated similarly. The only difference between them is when the operations are applied in the circuit. State preparation errors will occur after the initial state of the system is prepared, and measurement errors occur when the ancilla qubits are measured and when a final measurement is made on the qubit system.
Since SPAM instruments are not exact, it is possible to initialize or measure the incorrect state. In this model, two gates are applied on this assumption.

$$ K_{M_0} = \sqrt{1-P_2}I, \quad K_{M_1} = \sqrt{P_2}X$$

where $P_2$ is the probability for error. This value can be different for state preparation and measurement errors respectively. Thus the effect of the state preparation and measurement channel on the density matrix $\rho$ in this case can be defined as 

$$\rho \mapsto S(\rho) = K_{M_0}\rho K_{M_0} + K_{M_1}\rho K_{M_1}$$

Although it can be done similarly to the depolarization error model, \coolname does not currently offer a qubit-by-qubit variation on SPAM error rates.

\section{\label{sec:features}General features of quantum error correcting codes}

To successfully implement quantum error correcting codes, there are a few requirements and ideas that must be kept in mind. Firstly it is impossible to perfectly copy an unknown quantum state, and second, you cannot directly measure a quantum state without destroying it \cite{Devitt_2013}. Thus, when implementing and/or creating QEC codes it is useful to encode multiple physical qubits into a single logical qubit. When done correctly, with a larger Hilbert space it is possible to limit the codeword states (i.e. $\vert000\rangle \equiv \vert0\rangle_L$ when encoding 3 qubits as a single logical qubit). This allows errors on individual qubits to be identified using ancilla (syndrome) qubits. This is possible due to the limited codespace, which is a subspace of the larger Hilbert space where the codeword states live. When a state exists outside this subspace, the ancilla qubits will be able to detect the errors that occurred, and a correction can be made.

In this work, quantum gates are needed to initialize codeword states, detect quantum errors, and correct for quantum errors. Further reading can be done to understand quantum gates and operations including section 4 of \cite{10.21468/SciPostPhysLectNotes.70} and \texttt{01.Introduction to Quantum Error.ipynb}.

\subsection{\label{sec:connectivity}Qubit connectivities}

When implementing quantum circuits, it is important to limit the number of gates that are implemented and pay attention to the qubits that are used for implementing certain gates. One limitation of superconducting quantum circuits is that once they are fabricated, the layout may not be changed. This means that when implementing quantum circuits with gate operations, the qubit connectivity will affect how many gates are needed. This is because applying gates between non-adjacent qubits will require subsequent gates to take place between multiple qubits, greatly increasing gate count and diminishing circuit fidelity. This is where qubit connectivities are important to think about.

\coolname implements 3 types of connectivity: all-to-all, line, and square lattice. Having an all-to-all connected circuit means that every qubit-qubit gate operation only requires a single operation to take place, which is the ideal case. For some QEC codes, it is possible to physically implement all-to-all, however, it is not realistically feasible for most. In most superconducting qubit systems, groups focus on line-connected and square lattice-connected systems as they can be easier to fabricate. Being limited by connectivity can greatly impact circuit fidelity. This is due to an increase in required circuit depth and time, leading to higher error rates. In addition to number of gates applied, where these gates are applied also matters. As different qubits have different parameters and some gates may be more accurate if implemented on only certain qubits. 

In this work, qubits were placed in line-connected circuits as shown in Figure \ref{fig:threequbitcode} where each qubit is connected only to its adjacent qubit(s), and no rigorous optimizations were made (we attempted efficient implementations but do not claim our implementation is provably optimal).
For the square lattice-connected circuits qubits were placed along a square grid such that most gates were easily accessible between qubits.
Again, we make no rigorous claims about optimization in terms of minimizing the number of gates and best qubit-gate parameters.

\subsection{Implementing non-adjacent quantum gates with \coolname}
To implement CNOT gates between non-adjacent qubits in a line-connected or square lattice-connected circuit, this work follows derivations in \cite{Bataille2020}.
Similar ideas are also used to implement CZ gates.
When creating quantum circuits, \coolname can determine whether to use adjacent or non-adjacent gates (for CNOT and CZ) depending on the parameters given by the user. 
Furthermore, if certain qubit connectivity is required, \coolname offers the flexibility of choosing adjacent or non-adjacent gate operations by the user. 

\section{\label{sec:codes}Codes in \coolname} 



The codes available in \coolname were selected based on the following requirements: simplicity and low qubit count.
The goal with these codes is not to construct a fault-tolerant quantum computer, but to offer a diagnostic tool for teams working with of order one dozen qubits with the primary goal of better understanding noise and qubit performance characteristics by comparing hardware runs to simulation results. Most of the implementation details below follow \cite{Devitt_2013}.

\subsection{\label{sec:threequbit}The 3-qubit Code}
The most basic error-correcting code is the 3-qubit code as it requires the least number of physical qubits and gate operations. This error-correcting code, shown in Figure \ref{fig:threequbitcode} can correct for a maximum of one bit flip (or phase flip depending on how the code is set up) error on a single data qubit in the system. If any more than a single error occurs the error correction fails. Figure \ref{fig:threequbitcode} shows the final location where the error, $\mathcal{E}$, can occur without causing the circuit to fail. Furthermore, if an error occurs on one of the ancilla qubits, it is most dangerous due to errors allowed to spread throughout the system by applying an inaccurate correction. 

\subsubsection{Logical state initialization}
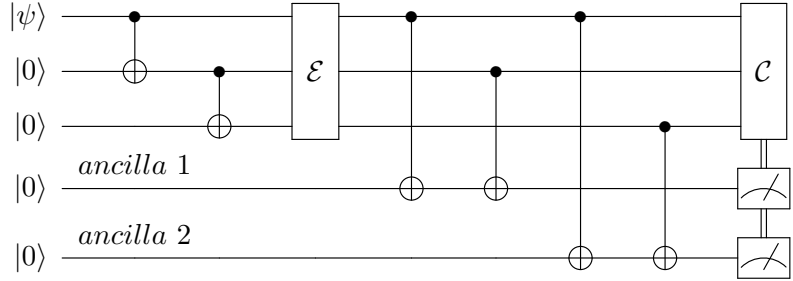
\begin{figure}
\[
\begin{array}{c}
\Qcircuit @C=2.1em @R=1em {
\lstick{\ket{\psi}} & \ctrl{1}               & \qw      & \multigate{2}{\mathcal{E}} & \ctrl{3} & \qw      & \ctrl{4} & \qw      & \multigate{2}{\mathcal{C}} \\
\lstick{\ket{0}}    & \targ                  & \ctrl{1} & \ghost{\mathcal{E}}        & \qw      & \ctrl{2} & \qw      & \qw      & \ghost{\mathcal{C}} \\
\lstick{\ket{0}}    & \qw                    & \targ    & \ghost{\mathcal{E}}        & \qw      & \qw      & \qw      & \ctrl{2} & \ghost{\mathcal{C}} \cwx[1] \\
\lstick{\ket{0}}    & \ustick{ancilla~1} \qw & \qw      & \qw                        & \targ    & \targ    & \qw      & \qw      & \meter \cwx[1] \\
\lstick{\ket{0}}    & \ustick{ancilla~2} \qw & \qw      & \qw                        & \qw      & \qw      & \targ    & \targ    & \meter \\
}
\end{array}
\]
\caption{
The 3-qubit code.
Here $\mathcal{E}$ denotes where an error occurs (such that the code can correct it), and $\mathcal{C}$ is the correction, based on ancilla measurements.
}
\label{fig:threequbitcode}
\end{figure}

The first step of a quantum error correction code is to initialize the logical state to be in the codespace. In the 3-qubit code this is done with 2 CNOT gate operations between the 3 data qubits. In Figure \ref{fig:threequbitcode}, $\vert\psi\rangle$ is a single qubit state that is encoded into the logical qubit. During the initialization, the system state is transformed from $\vert\psi\rangle\vert00\rangle \mapsto \vert\psi\psi\psi\rangle \equiv \vert\psi\rangle_L.$

After initialization, quantum errors, $\mathcal{E}$ can be applied as shown in figure \ref{fig:threequbitcode}. It is important to note that realistically, and in error models implemented, quantum errors occur randomly and throughout the entire circuit process. The circuit in Figure \ref{fig:threequbitcode} is used to show the final location where a discrete single bit flip error can be accurately detected and finally corrected.

\subsubsection{Error detection and correction}
To detect a bit flip error that occurs on one of the data qubits, ancilla qubits are introduced. These qubits are initialized to $\vert0\rangle$ and coupled to the system. A CNOT gate is applied between data qubits and ancilla qubits such that a parity check is made between two data qubits. Since there are three data qubits in the system, it is sufficient to apply checks between just two pairs of data qubits. The parity check information is stored in the ancilla qubits and they are measured. Depending on their state ($\vert00\rangle$, $\vert01\rangle$, $\vert10\rangle$, or $\vert11\rangle$) it is possible to determine the location of the bit flip error that occurred in the data qubits. The correction is then made by applying a $\sigma_X$ gate on the qubit where the error was detected.

\subsubsection{Logical state failure}
To properly benchmark various QEC algorithms such as the three-qubit code, the nine-qubit code, and the Steane code, it is important to know when circuit failure occurs. When a QEC circuit fails, the logical state of the qubit system is no longer measured to be the initial state. This means that the rate at which errors occurred caused the correction algorithm to fail.

This analysis can be done by repetitively running the specific QEC code that is being tested. After each repetition of the algorithm, a measurement of the qubit system is made. If the qubit state remains in the initial state, then the circuit is run again. However, if the qubit state is not measured to be in the initial state, the circuit stops running. This process is repeated many times for each circuit and the number of iterations until failure is counted for each of these repetitions.

\begin{figure}
\includegraphics[width=150mm,scale=1]{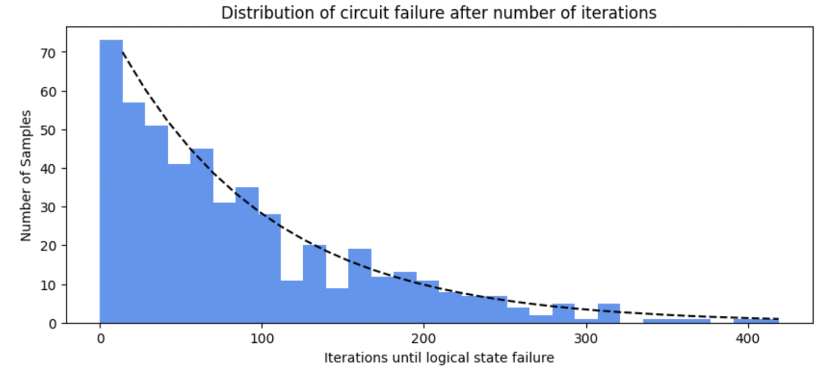}
\caption{\textbf{Three-Qubit Code Circuit Failure Distribution.} The distribution between the number of samples sorted by the number of iterations that the three-qubit code runs before failing is shown. In this case, each qubit is set to have physical parameters similar to current realized devices. The three-qubit QEC code on current qubits is unsuccessful, due to the large number of sample failures before reaching more iterations.}
\label{fig:three_qubit_hist}
\end{figure}

This distribution for the three-qubit code is shown in Figure \ref{fig:three_qubit_hist}. After analyzing the distribution for circuit failures, it is clear that the three-qubit QEC code is unsuccessful in maintaining the correct logical state with current qubit performance levels. This makes sense because the three-qubit code can only correct for a maximum of one bit error per iteration. Thus if more occur, the circuit will not be able to distinguish these errors. Since this QEC code only utilizes 5 physical qubits, the computational memory and time needed are limited so it is possible to run the circuit many times. On the other hand, for the 9-qubit code (utilizes 11 physical qubits) and the Steane code (utilizes 10 physical qubits), the memory and time needed to run the circuit exceed a laptop's abilities. Further work is needed to continue the analysis for other circuits, thus this subsection will be omitted in the Steane code and nine-qubit code discussions.

\subsection{\label{sec:steane}The Steane Code}
The Steane code, shown in Figure \ref{fig:steanecode}, is one of the most well known quantum error correction codes and is a $[[n, k, d]] = [[7, 1, 3]]$ quantum code. This means that it encodes $n = 7$ physical qubits into $k = 1$ logical qubits and has a distance  of $d = 3$ between basis states. Thus that it allows up to 3 individual qubit operations to maintain the same basis state, any more and the basis state will change. From this it is possible to calculate the number of errors that the Steane code can accurately correct, which is $\left\lfloor(d-1)/2\right\rfloor = 1$ error (either bit flip or phase flip) on any data qubit.

\subsubsection{Stabilizer formalism}
When implementing the 3-qubit code, parity checks were made using 2 ancilla qubits. The gate operations made on the system were were described using the state vector representation. When implementing smaller codes where the state and circuit is not changing from run to run, this representation is sufficient. However, stabilizer formalism can be more efficient for representing quantum error correction codes. \texttt{03.Stabilizer Codes and Steane Code.ipynb} section A, demonstrates the use of stabilizer formalism to implement the 3-qubit code as a useful example. 

Stabilizer formalism takes the larger Hilbert space and provides constraints to the system using stabilizer operators such that the data qubits are constrained to a smaller subspace. This subspace is called the codespace and, if the system state is found outside of the codespace, an error occurred and a correction can be applied. This makes it easier to see what types of logical operations can be made to encode data and is relatively easy to implement. The two codeword states that are in the codespace for the Steane code are shown in Equation \ref{eq:steanecodewords}.

\begin{equation}
\begin{aligned}
\vert0\rangle_L = \frac{1}{\sqrt{8}}(\vert0000000\rangle + \vert1010101\rangle + \vert0110011\rangle + \vert1100110\rangle \\
+ \vert0001111\rangle + \vert1011010\rangle + \vert0111100\rangle + \vert1101001\rangle)\\
\vert1\rangle_L = \frac{1}{\sqrt{8}}(\vert1111111\rangle + \vert0101010\rangle + \vert1001100\rangle + \vert0011001\rangle \\
+ \vert1110000\rangle + \vert0100101\rangle + \vert1000011\rangle + \vert0010110\rangle)
\end{aligned}
\label{eq:steanecodewords}
\end{equation}

\subsubsection{Logical State initialization}
In this work, the Steane code is implemented using stabilizer formalism. For a n qubit state the dimension of the Hilbert space is $2^n$, however for a single logical qubit (and to  we must restrict to a 2 dimensional Hilbert space. The stabilizer used for the 7-qubit logical states above, $\ket{0}_L$ and $\ket{1}_L$, consist of the six operators in Equation \ref{eq:steanestabilizers}. The final operator ($Z^{\otimes 7}$) fixes the state into one of the two codewords.

\begin{equation}
\begin{split}
K^1 =&~ IIIXXXX \\
K^2 =&~ XIXIXIX \\
K^3 =&~ IXXIIXX \\
K^4 =&~ IIIZZZZ \\
K^5 =&~ ZIZIZIZ \\
K^6 =&~ IZZIIZZ \\
\end{split}
\label{eq:steanestabilizers}
\end{equation}

\begin{figure}
\[
\begin{array}{c}
\Qcircuit @C=2.1em @R=1.75em {
\lstick{\ket{0}} & \qw            & \gate{H} & \qw                            & \qw                            & \ctrl{3}                       & \gate{H} & \meter \cwx[1]                 & M_3 & \\
\lstick{\ket{0}} & \qw            & \gate{H} & \qw                            & \ctrl{2}                       & \qw                            & \gate{H} & \meter \cwx[1]                 & M_2 & \\
\lstick{\ket{0}} & \qw            & \gate{H} & \ctrl{1}                       & \qw                            & \qw                            & \gate{H} & \meter \cwx[1]                 & M_1 & \\
\lstick{\ket{0}} & \ustick{7} \qw & \qw      & \multigate{6}{\mathcal{K}^{1}} & \multigate{6}{\mathcal{K}^{2}} & \multigate{6}{\mathcal{K}^{3}} & \qw      & \multigate{6}{\mathcal{Z}_{i}} & \qw & \\
\lstick{\ket{0}} & \ustick{6} \qw & \qw      & \ghost{\mathcal{K}^{1}}        & \ghost{\mathcal{K}^{2}}        & \ghost{\mathcal{K}^{3}}        & \qw      & \ghost{\mathcal{Z}_{i}}        & \qw & \\
\lstick{\ket{0}} & \ustick{5} \qw & \qw      & \ghost{\mathcal{K}^{1}}        & \ghost{\mathcal{K}^{2}}        & \ghost{\mathcal{K}^{3}}        & \qw      & \ghost{\mathcal{Z}_{i}}        & \qw & \\
\lstick{\ket{0}} & \ustick{4} \qw & \qw      & \ghost{\mathcal{K}^{1}}        & \ghost{\mathcal{K}^{2}}        & \ghost{\mathcal{K}^{3}}        & \qw      & \ghost{\mathcal{Z}_{i}}        & \qw & \ket{0_L} \\
\lstick{\ket{0}} & \ustick{3} \qw & \qw      & \ghost{\mathcal{K}^{1}}        & \ghost{\mathcal{K}^{2}}        & \ghost{\mathcal{K}^{3}}        & \qw      & \ghost{\mathcal{Z}_{i}}        & \qw & \\
\lstick{\ket{0}} & \ustick{2} \qw & \qw      & \ghost{\mathcal{K}^{1}}        & \ghost{\mathcal{K}^{2}}        & \ghost{\mathcal{K}^{3}}        & \qw      & \ghost{\mathcal{Z}_{i}}        & \qw & \\
\lstick{\ket{0}} & \ustick{1} \qw & \qw      & \ghost{\mathcal{K}^{1}}        & \ghost{\mathcal{K}^{2}}        & \ghost{\mathcal{K}^{3}}        & \qw      & \ghost{\mathcal{Z}_{i}}        & \qw & \\
}
\end{array}
\]
\caption{
The Steane code initialization.
Here the $\mathcal{K}^{1,2,3}$ denote the appropriate multi-qubit stabilizer, and $\mathcal{Z}_i$ is the correction, based on the three ancilla measurements, where $i = 1^{M_2} + 2^{M_3} + 4^{M_1}$.
}
\label{fig:steanecode}
\end{figure}
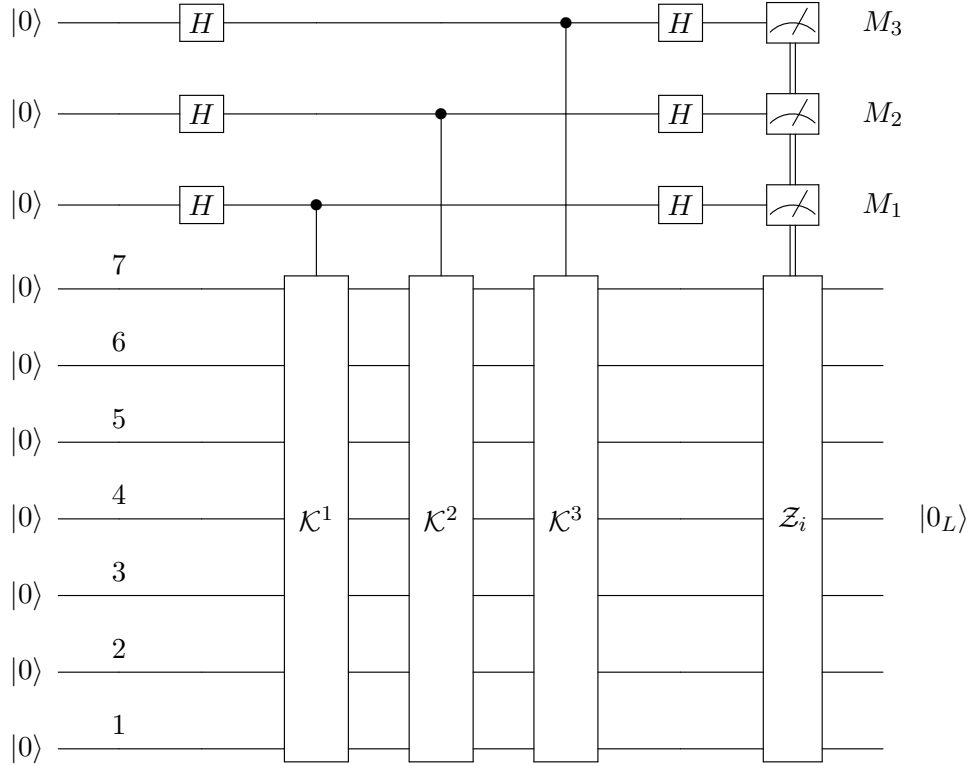

Figure \ref{fig:steanecode} shows the circuit that this used to initialize 7 qubits into the $\vert0\rangle_L$ state, and it uses 3 ancilla qubits to achieve this. Each ancilla qubit is initialized with a Hadamard gate and then is used as a control qubit for a stabilizer operator. This will project the initial state to eigenstates of each of the X stabilizer operators from Equation \ref{eq:steanestabilizers} ($K^1$, $K^2$, $K^3$).

A Hadamard gate is applied again to each ancilla qubit. This operation will change the state of each syndrome ancilla qubits if the 7 data qubits are not in the codespace. If this occurs, we apply a single qubit $Z_i$ gate depending on what the syndrome measurement is, where $i$ is one of the data qubits. The same process is repeated for the Z stabilizer ($K^4$, $K^5$, $K^6$) in Equation \ref{eq:steanestabilizers}, however the correction is $X_i$. When performing stabilizer operators, the system is put into a $+1$ eigenstate of the stabilizer. If there is an error, then it would be in a $-1$ eigenstate, which would cause the ancilla measurement to detect this. 

\subsubsection{Error detection and correction}
The process for detecting and correcting errors using the Steane code is the same as the initialization process. This is because if an error occurs on one of the qubits, and it commutes with the stabilizer, then the state remains in a $+1$ eigenstate of the stabilizer. However if the error does not commute, then the state will be a $-1$ eigenstate of the stabilizer and the same procedure can be used to correct it. More detailed information can be found in \cite{Devitt_2013} and \texttt{03.Stabilizer Codes and Steane Code.ipynb}.

\subsection{\label{sec:ftsteane} Fault-tolerance and The Steane Code}

In previous circuits, ancilla qubits were not protected from any type of error. Although they are used to detect and (hopefully) correct for errors, ancilla qubits can be the most detrimental to the fidelity of a quantum circuit. If an error occurs on a data qubit, a correction algorithm can be in place to help correct it, and if multiple data errors occur, there is still a chance for correction. However, if an ancilla qubit experiences an error, incorrect information can be propagated throughout the circuit. This causes the rate of errors to increase exponentially.

Take this example: Imagine in an implementation of the 3 qubit code our CNOT gates are correct only 80$\%$ of the time. There are 5 CNOT gates in the 3 qubit code, so there is a good chance that at least one of the gates will be incorrectly applied. Thus when the ancilla qubits are measured the syndrome correction would be incorrect. This is also assuming an all-to-all connectivity, in reality, we have many more CNOT gates that are being applied so our error count will increase.

Fault tolerance is the ability to stop these errors from propagating through our system. A fault-tolerant circuit will only allow an error to cause at most a single error in the output for each logical qubit block. Errors are bound to occur, but as long as they are stopped before they cascade, an error correction protocol can be applied. One important detail is that the definition of fault tolerance can change from system to system. If the error-correcting code can fix multiple errors then the definition above can be relaxed. Most of the time, it is enough to say a system that corrects for k errors is fault tolerant if 1, 2, ..., or k errors cause a maximum of k errors to occur in a logical qubit block output.

\subsubsection{The (Fault-Tolerant) Steane Code}
In \coolname, the Steane code is also implemented fault tolerantly to compare it to the "normal" Steane code. Here, only the propagation of a single error during the preparation is considered, since any more than one error is outside the capabilities of the Steane code. To construct a fault-tolerant state preparation circuit, the stabilizer operators must be measured in a fault-tolerant way. There are many ways to do this, but in this case, \cite{Devitt_2013} is followed.

To set up the circuit, four ancilla qubits (rather than three like previously) are used and are prepared in the state

$$\frac{1}{\sqrt2} (\vert0000\rangle + \vert1111\rangle).$$

This state can be reached by applying a Hadamard to one of the ancillas and then implementing a set of CNOT gates between them. Each ancilla will be used to control a single CNOT gate operation on four qubits (applying the stabilizer operator), after which the ancilla block is decoded and measured. This will ensure that any $\hat{X}$ error will only propagate to a single qubit in the data block, which can later be corrected. In addition to extra ancilla qubits, certain checks are made to this set of qubits. If these checks are not satisfied then the ancilla initialization is repeated. Furthermore, throughout the Steane code, the full ancilla detection protocol is iterated three times for each syndrome measurement, and the result that has a majority is used as the detection result. This ensures that error propagation is minimized and the code can run smoothly.

Due to the use of a single logical ancilla qubit, the Steane code is run sequentially where each stabilizer measurement is done one at a time, saving each of the results. These results are compared with one another and a majority vote is made on the measured error syndrome. This increases the number of operations needed to complete the circuit, thus exponentially increasing the computational resources necessary for simulating the circuit in \coolname. For a more detailed description of the Fault Tolerant Steane code see \texttt{04a.Fault Tolerant Steane Code.ipynb}.

\subsection{\label{sec:ninequbit}The 9-qubit Code}

The 9-qubit code is very similar to the 3-qubit code since it is also a repetition code and it even uses 3 sets of the 3-qubit code in its construction. However, with the use of 9 data qubits (adding ancilla qubits makes the total 11), the strength of the code increases. The 9-qubit code can correct for up to three bit flip ($\hat{X}$) errors and one phase error ($\hat{Z}$). Due to the high similarities, only the major differences will be discussed in this section.

The key difference is the use of 9 qubits rather than just 3 allowing for the code to detect phase errors. These errors are detected due to the addition of Hadamard and CNOT gates applied to the 11-qubit system. With this, there is an increased number of gate operations, thus a realistic error model will lead to faster code failure in terms of iterations. The reason iterations are specified rather than time is that each gate takes the same amount of time, thus with more gates, that doesn't necessarily decrease circuit time, but rather full circuit iterations.
\begin{figure}
\[
\begin{array}{c}
\Qcircuit @C=2.1em @R=1em {
\lstick{\ket{\psi}} & \ctrl{3} & \ctrl{6} & \gate{H} & \ctrl{1} & \ctrl{2} & \qw & \\
\lstick{\ket{0}}    & \qw      & \qw      & \qw      & \targ    & \qw      & \qw & \\
\lstick{\ket{0}}    & \qw      & \qw      & \qw      & \qw      & \targ    & \qw & \\
\lstick{\ket{0}}    & \targ    & \qw      & \gate{H} & \ctrl{1} & \ctrl{2} & \qw & \\
\lstick{\ket{0}}    & \qw      & \qw      & \qw      & \targ    & \qw      & \qw & \ket{\psi_L}\\
\lstick{\ket{0}}    & \qw      & \qw      & \qw      & \qw      & \targ    & \qw & \\
\lstick{\ket{0}}    & \qw      & \targ    & \gate{H} & \ctrl{1} & \ctrl{2} & \qw & \\
\lstick{\ket{0}}    & \qw      & \qw      & \qw      & \targ    & \qw      & \qw & \\
\lstick{\ket{0}}    & \qw      & \qw      & \qw      & \qw      & \targ    & \qw & \\
}
\end{array}
\]
\caption{
The circuit required to encode a single logical qubit using Shor's 9-qubit code.
}
\label{fig:ninequbitcode}
\end{figure}
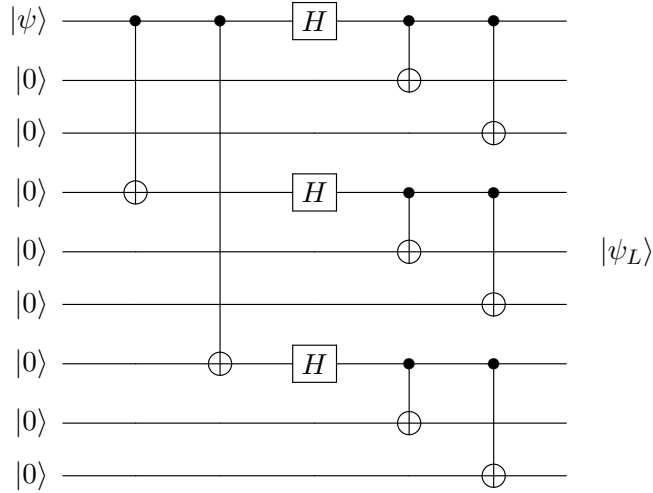
By encoding our 9 data qubits into two logical states ($\vert0\rangle_L$ and $\vert1\rangle_L$), this QEC circuit works similarly to others. The circuit initialization is shown in Figure \ref{fig:ninequbitcode}, and the two logical states are shown below.

$$ \vert0\rangle _L = \frac{1}{\sqrt{8}}(\vert000\rangle + \vert111\rangle)(\vert000\rangle + \vert111\rangle)(\vert000\rangle + \vert111\rangle) $$
$$ \vert1\rangle _L = \frac{1}{\sqrt{8}}(\vert000\rangle - \vert111\rangle)(\vert000\rangle - \vert111\rangle)(\vert000\rangle - \vert111\rangle) $$

\begin{figure}
\[
\begin{array}{c}
\Qcircuit @C=2.1em @R=1em {
\lstick{\ket{\psi}} & \gate{H}           & \ctrl{9} & \qw      & \qw      & \qw      & \qw      & \qw      & \qw      & \qw      & \qw      & \qw      & \qw      & \qw      & \gate{H} & \\
\lstick{\ket{0}}    & \gate{H}           & \qw      & \ctrl{8} & \qw      & \qw      & \qw      & \qw      & \qw      & \qw      & \qw      & \qw      & \qw      & \qw      & \gate{H} & \\
\lstick{\ket{0}}    & \gate{H}           & \qw      & \qw      & \ctrl{7} & \qw      & \qw      & \qw      & \qw      & \qw      & \qw      & \qw      & \qw      & \qw      & \gate{H} & \\
\lstick{\ket{0}}    & \gate{H}           & \qw      & \qw      & \qw      & \ctrl{6} & \qw      & \qw      & \ctrl{7} & \qw      & \qw      & \qw      & \qw      & \qw      & \gate{H} & \\
\lstick{\ket{0}}    & \gate{H}           & \qw      & \qw      & \qw      & \qw      & \ctrl{5} & \qw      & \qw      & \ctrl{6} & \qw      & \qw      & \qw      & \qw      & \gate{H} & \\
\lstick{\ket{0}}    & \gate{H}           & \qw      & \qw      & \qw      & \qw      & \qw      & \ctrl{4} & \qw      & \qw      & \ctrl{5} & \qw      & \qw      & \qw      & \gate{H} & \\
\lstick{\ket{0}}    & \gate{H}           & \qw      & \qw      & \qw      & \qw      & \qw      & \qw      & \qw      & \qw      & \qw      & \ctrl{4} & \qw      & \qw      & \gate{H} & \\
\lstick{\ket{0}}    & \gate{H}           & \qw      & \qw      & \qw      & \qw      & \qw      & \qw      & \qw      & \qw      & \qw      & \qw      & \ctrl{3} & \qw      & \gate{H} & \\
\lstick{\ket{0}}    & \gate{H}           & \qw      & \qw      & \qw      & \qw      & \qw      & \qw      & \qw      & \qw      & \qw      & \qw      & \qw      & \ctrl{2} & \gate{H} & \\
\lstick{\ket{0}}    & \ustick{ancilla~1} & \targ    & \targ    & \targ    & \targ    & \targ    & \targ    & \qw      & \qw      & \qw      & \qw      & \qw      & \qw      & \qw & \\
\lstick{\ket{0}}    & \ustick{ancilla~2} & \qw      & \qw      & \qw      & \qw      & \qw      & \qw      & \targ    & \targ    & \targ    & \targ    & \targ    & \targ    & \qw & \\
}
\end{array}
\]
\caption{
The circuit required to detect a phase error ($\hat{Z}$) in the Shor's 9-qubit code. The first six CNOT gates compare the phase of the first block with that of the second block, and the next set of six CNOT gates compare the second block with the third. The ancilla qubits are measured after the process is finished and a phase correction is applied to one of the qubits in the block with the incorrect phase.
}
\label{fig:ninequbitcode_phase}
\end{figure}
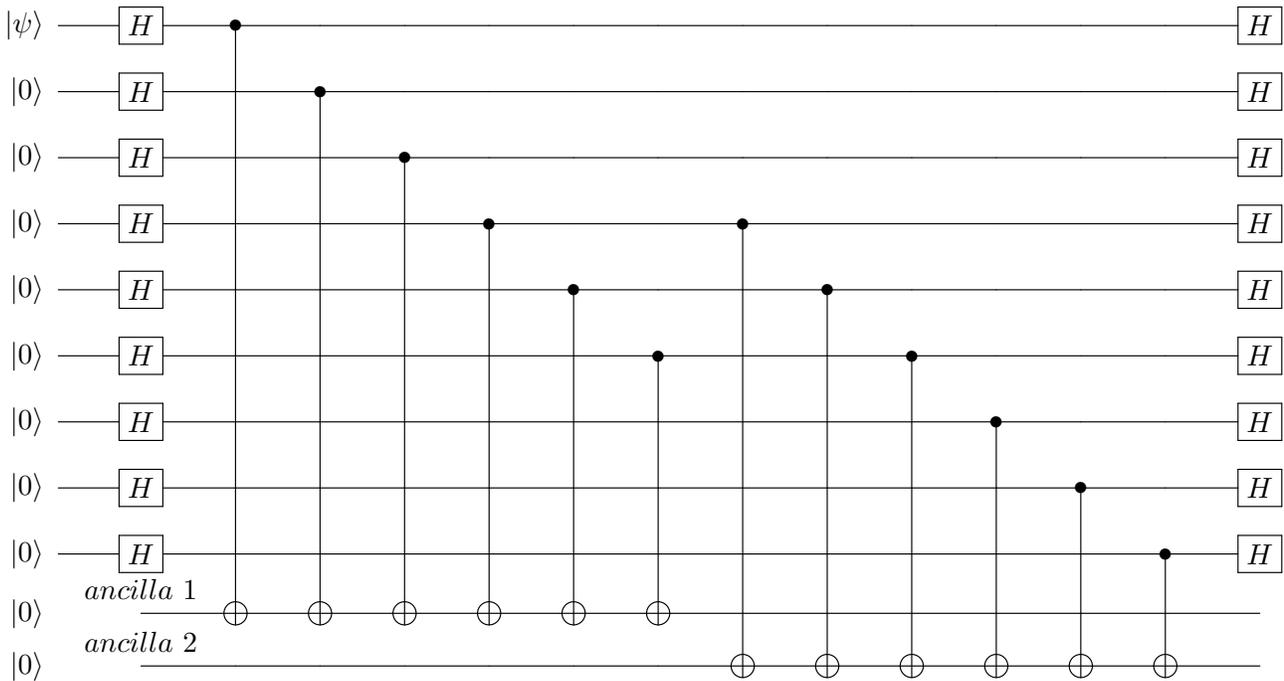

The 9-qubit code sequentially corrects for the phase and then bit error(s) in the circuit. First, the phase error is corrected by applying a Hadamard gate on each of the data qubits. Then a phase comparison is done between each of the three blocks using CNOT gates, and the comparison is stored in the ancilla. Finally, a Hadamard is applied again on the data qubits and a correction can be made on one of the three blocks. Here, only a single $\sigma_Z$ gate is applied on one of the qubits of the "faulty" data block, which, is enough to correct for the phase error of the whole block. The phase error detection code can be seen in Figure \ref{fig:ninequbitcode_phase}.

The error detection process for the bit flip error works exactly like the 3-qubit code. The same detection code is iterated three times (one for each block of three) on the 9 qubits. During each iteration the three qubits in the set are coupled to the two ancilla qubits and a single bit flip error is corrected. Then the ancillas are reset and the process repeats for each of the blocks.

\section{\label{sec:outlook}Future work}

FTQC is not likely to use any of the codes currently implemented in \coolname.
This is because these codes are extremely sensitive to qubit noise.
Probably the most studied code for FTQC is the surface code \cite{PhysRevA.86.032324,Tomita2014,Fowler2014}, primarily due to its ability to function under much higher local error rates. 
The surface code is also capable of handling relaxation errors, which is something the currently implemented codes cannot do.
Using surface codes, different groups have been able to demonstrate QEC featuring performance improvements scaling with higher qubit counts \cite{Acharya2023}.
The surface code is not an immediate priority for \coolname because realistic implementations require several physical qubits far beyond the desktop scale. 
As such, the surface code is somewhat less friendly as a tool for studying qubit noise and performance.

An additional, related possible extension would be to allow for multiple logical qubits and include features like magic state distillation, etc. \cite{Bravyi_2005}
This would likely require working at the logical qubit level with physical qubits largely abstracted away or simulation costs would be prohibitive for studying any realistic code for fault-tolerant computation.

Likely the most important missing ingredient from the noise model at this point is \emph{cross-talk}, which is the noise created on a physical qubit when addressing a different physical qubit by the coupling between them or the address method, see e.g. \cite{PhysRevA.101.052308}.
One of the authors has prior work on a cross-talk model that we may eventually port into \coolname \cite{gustafson2021large}.

\coolname's usefulness is limited by the long wall-clock time required to run multiple repetitions of codes that utilize larger qubit counts.
This situation derives from state vectors and operators with memory footprints that grow exponentially with system size, forcing the classical computer simulating the quantum system to swap memory during computations.
The library currently makes no effort to minimize the memory footprint of computations, for example, by leveraging sparse matrices, and this is a priority for future development, particularly where drop-in replacements may be an option.

Currently, the noise model in \coolname is ``global-uniform,'' by which we mean that specified parameters, e.g. T1, are the \emph{same} for every physical qubit in the logical qubit. 
Of course, on modern quantum hardware, this is never the case.
Furthermore, it does not make sense to always assume that utilizing the worst-case features on a physical QPU will accurately represent the final performance because it is often possible to design logical qubit layouts in a ``noise aware'' fashion that minimizes interactions with the worst performing pieces of the hardware.
Therefore, implementing qubit and coupler-specific noise parameters efficiently is a priority.

\section{\label{sec:conclusion}Conclusions}


In this paper we presented \coolname, an open-source, portable, easily extended toolkit for studying qubit noise characteristics in the context of QEC codes.
The library features a functional approach and a shallow function hierarchy to emphasize pedagogical use cases and ease of reading and modification.
The package includes a user-friendly notebook for estimating code performance (in terms of repetition cycles) for a parameterized noise model as well as a set of notebooks illustrating all the concepts employed in the software.

It is our hope that groups studying the performance of quantum processors with order one dozen qubits find \coolname useful for performance benchmarking and experiment design.
\coolname is an open source research tool and the authors encourage contributions and modifications.

\section{\label{sec:ack}Acknowledgements}

\coolname was developed primarily by J.A.P. and S.L. during the Summer of 2023 as a student project in the Open Quantum Initiative (OQI) Undergraduate Fellowship\footnote{\url{https://chicagoquantum.org/oqi-undergraduate-fellowship}} program.
The authors would all like to thank the OQI program managers, staff, and other participants for their support.

J.A.P. and S.L. were financially supported by Practices to Accelerate the Commercialization of Technologies (PACT).
G. P. was financially supported for this work by the DOE/HEP QuantISED program grant ``HEP Machine Learning and Optimization Go Quantum,'' identification number 0000240323.

This document was prepared using the resources of the Fermi National Accelerator Laboratory (Fermilab), a U.S. Department of Energy (DOE), Office of Science, HEP User Facility. 
Fermilab is managed by Fermi Research Alliance, LLC (FRA), acting under Contract No. DE-AC02-07CH11359.

\newpage

\bibliographystyle{unsrt}
\bibliography{references}

\end{document}